\documentclass[a4paper,10pt,twocolumn]{article}

\usepackage[utf8]{inputenc}
\usepackage{multicol}
\usepackage{graphicx}
\usepackage{times}
\usepackage{hyperref}
\usepackage[T1]{fontenc}

\usepackage[backend=bibtex,giveninits=true, doi=false,isbn=false,url=false,eprint=false]{biblatex}

\defbibenvironment{bibliography}
  {\list 
    {\printfield[labelnumberwidth]{labelnumber}} 
  {\setlength{\labelwidth}{0pt}%
   \setlength{\leftmargin}{0pt}%
   \setlength{\labelsep}{4pt}%
   \setlength{\itemsep}{\bibitemsep}%
   \setlength{\itemindent}{\labelwidth}%
   \addtolength{\itemindent}{\labelsep}%
   \setlength{\parsep}{\bibparsep}}%
   } 
{\endlist} 
{\item}
\DeclareNameAlias{sortname}{last-first}
\DeclareNameAlias{author}{last-first}

\bibliography{biblio}

\fontfamily{ptm}\selectfont

\usepackage[left=2cm,right=2cm,top=2.5cm,bottom=2.5cm]{geometry}

\usepackage{titlesec}
\titleformat*{\section}{\MakeUppercase}

\begin{document}

\title{The Impact of Flow in an EEG-based Brain Computer Interface}

\twocolumn[{
\begin{center}
	\begin{Large}
\textbf{The Impact of Flow in an EEG-based Brain Computer Interface}\\
	\end{Large}
    \begin{large}
        \vspace{0.6cm}
        J. Mladenovi\'c$^{1,2}$, J. Frey$^{1,3}$, M. Bonnet-Save$^{1}$, J. Mattout$^{2}$, F. Lotte$^{1}$\\
        \vspace{0.6cm}
        $^{1}$Inria, Bordeaux, France\\
        $^{2}$INSERM U1028, Lyon, France \\
        $^{3}$Ullo, La Rochelle, France\\
        \vspace{0.5cm}
        E-mail: jelena.mladenovic@inria.fr
        \vspace{0.4cm}
    \end{large}
\end{center}
}]

ABSTRACT:
Major issues in Brain Computer Interfaces (BCIs) include low usability and poor user performance.
This paper tackles them by ensuring the users to be in a state of immersion, control and motivation, called state of flow. Indeed, in various disciplines, being in the state of flow was shown to improve performances and learning. Hence, we intended to draw BCI users in a flow state to improve both their subjective experience and their performances. In a Motor Imagery BCI game, we manipulated flow in two ways: 1) by adapting the task difficulty and 2) by using background music. Results showed that the difficulty adaptation induced
a higher flow state, however music had no effect. There was a positive correlation between subjective flow scores and offline performance, although the flow factors had no effect (adaptation)
or negative effect (music) on online performance. Overall, favoring the flow state seems a promising approach for enhancing users’ satisfaction, although its complexity requires more thorough investigations.

\section*{Introduction}

The Brain Computer Interface (BCI) community today's priority is to assure the system robustness and its usability. It is quite a difficult task, considering the abundant inter and intra-subject variability. The major obstacle lies in the large spectrum of sources of variability during BCI usage, ranging from (i) imperfect recording conditions e.g. environmental noise, humidity, static electricity etc. \cite{Maby2016} to (ii) the fluctuations in the user’s psycho-physiological states, due to e.g., fatigue, motivation or attention \cite{JeunetPredictingPatterns}. There are yet more improvements to be done for a system ready to be easily used in real life conditions \cite{Wolpaw2002BraincomputerControl}. 

BCI systems showed quite an improvement with adaptive methods, i.e. adapting the machine to the changeable brain signals of the user during a BCI task. Currently, adaptation is mainly done by using different signal processing techniques without including human factors \cite{Makeig2012}. However, if the users do not understand how to manipulate a BCI system, or are not motivated to make necessary effort for such manipulation, then they are not able to produce stable and distinct EEG patterns. In that case, no signal processing algorithm would be able to decode such signals \cite{Lotte2013}. Thus, for designing a BCI, ignoring certain information about the users, e.g. their skills, cognitive abilities and motivations, may represent one of the major drawbacks for the advancement of BCIs.

A potential improvement in BCI is to acknowledge how difficult it can be to learn to produce mental commands (a very atypical skill) without a proper feedback about the progress one has made.
In every discipline, a certain feedback on ones performance is necessary to enable learning, as shown in the earliest work about Operant Conditioning and Reinforcement Learning \cite{Skinner1938}. Notably, this question was studied by behaviorists for decades on animals, using rewards e.g. food, as extrinsic motivation to promote desired behavior. As humans have more complex cognitive functions, a more effective way to promote learning is in a social context, with a tutor who would prepare and adapt a task according to the student's competences. The tutor's feedback and well organized tasks would lead the disciple to gradually build up knowledge and skills, to feel confident and to be intrinsically motivated, or to be in the Zone of Proximal Development (ZPD)\cite{Vygotsky1978}. Derived from cognitive developmental theories \cite{Vygotsky1978} and refined through instructional design theories \cite{Keller1987,Malone1987}, intrinsic motivation is to be a substantial element for learning. Thus, it is important to carefully design the feedback if we want to encourage learning and optimal performance. 

Unfortunately, for long this was not the case in BCI community, as BCI systems were improved mostly with novel machine learning techniques \cite{Makeig2012}. The result of neglecting the feedback design led to often monotonous and repetitive content, further discouraging the user, and leading to reduced skill and impaired performance \cite{Cho2004a,Kleih2010}, thus highly affecting the system's accuracy. Potentially, instructional design theories could add a missing piece for designing optimal BCI feedback \cite{Lotte2013}.

There have been extensive literature describing higher BCI user performance and experience using game-like feedback \cite{Ron-angevin2009BraincomputerTechniques,Scherer2015}. 
Immersive and game-like environments attract users' attention, induce intrinsic motivation, thus promote learning and performance with less effort and frustration -- for a review see \cite{Lumsden2016}.
Even using extrinsic motivation such as monetary reward can encourage users to perform better \cite{Kleih2010}. Some studies showed that user's belief on their performance with biased feedback induced motivation and thus higher performance \cite{Barbero2010BiasedInterfaces.}. Hence, sometimes it is worth to trade the system's accuracy to the perceived, subjective user's feeling of control.

Keeping that into account, a way to promote efficiency and motivation while respecting the principles of instructional design leads us to the Theory of Flow introduced by Csikszentmihalyi in \cite{Csikszentmihalyi1975}. He was fascinated by the capacity of artists to be in a state of enjoyment while effortlessly focused on a task so immersive that one looses the perception of time, of self and of basic human needs (hunger, sleep etc.). When in the flow state, people are absorbed in an activity, their focused awareness is narrowed, they lose self-consciousness, and they feel in control of their environment. As a consequence, they often perform to the best of their capacity with a sense of automaticity and a high level of confidence. Studies report flow experience in numerous activities including rock climbing, dancing, chess, reading, etc. \cite{Csikszentmihalyi1975,Csikszentmihalyi1989}. 

Another pertinent element which encourages intrinsic motivation and is showed to be in relation with flow, is music \cite{Croom2015}. Recent studies showed that music has an ergogenic effect on humans, i.e. physical enhancement while performing a physical activity \cite{Anshel1978}. In \cite{Karageorghis2010ErgogenicExercise} was reported that Haile Gebrselassie, an athlete who broke 10 000m world record in 1998, paced his running on music he was listening to, i.e. synchronous music. There is evidence that synchronous music, as a strong motivational effect, directly enhances physical performance \cite{Simpson2006} while asynchronous (background) music induces flow when accomplishing a task \cite{Pates2003EffectsPlayers,Pain2011Pre-CompetitionSoccer}. Most of all, background music with medium tempo (speed) has showed highest impact on flow \cite{Karageorghis2008PsychologicalExercise}.

To be in the state of flow, a task needs to have the following requirements:
\begin{itemize}

\item \textbf{To be immersive}, with attractive visual/audio stimuli to maintain the user's attention. The principle of preserving flow with aesthetically pleasing and ergonomic content have been researched largely in the context of human computer interaction \cite{Webster1993} and Internet navigation, e.g. e-learning \cite{Esteban-millat2014ModellingEnvironment};

\item \textbf{To adapt} the task difficulty with the user’s skills, i.e. an easy task might be boring as a difficult one might be frustrating, hence finding the golden middle is the way of feeling in control and keeping the motivation. Such difficulty adaptations were found in games, to keep the gamer in flow \cite{Bulitko2012FlowModel}, or during teaching activities \cite{Clement2015Multi-ArmedJEDM} to improve learning and keep the student in the ZPD \cite{Vygotsky1978}.

\item \textbf{To have clear goals and immediate feedback / rewards}; aspired for educational purposes \cite{Heutte2016}, so that learning becomes an enjoying and autotelic (self-rewarding) process \cite{Ninaus2015GameTask}.
\end{itemize}

Therefore, in order to improve BCI users' performance, learning, and experience, it seems promising to try to guide them towards the state of flow. This is what we start to explore in this paper.
In particular, our research question is: \textit{Does flow improve BCI user performance?} We chose to manipulate flow in a ludic BCI environment with 2 factors: 1) Feedback adaptation, i.e. perceived difficulty adaptation to the user skills, and 2) Asynchronous music to encourage the user. Thus, our following hypotheses are:

\textbf{H1.} Adapting the feedback improves flow, thus improves performance. 

\textbf{H2.} Asynchronous music improves flow, thus improves performance.

In consonance with the Flow theory, we presented a motor imagery (MI) BCI task in an open-source 3D video game (TuxRacer \footnote{https://extremetuxracer.sourceforge.io/}).  
We investigated the effects of these two flow factors on user's flow state as well as on user performance, i.e. classification accuracy.

\section*{Materials and Methods}

\begin{figure}[!htbp]
    \centering
    \includegraphics[width=0.7\columnwidth]{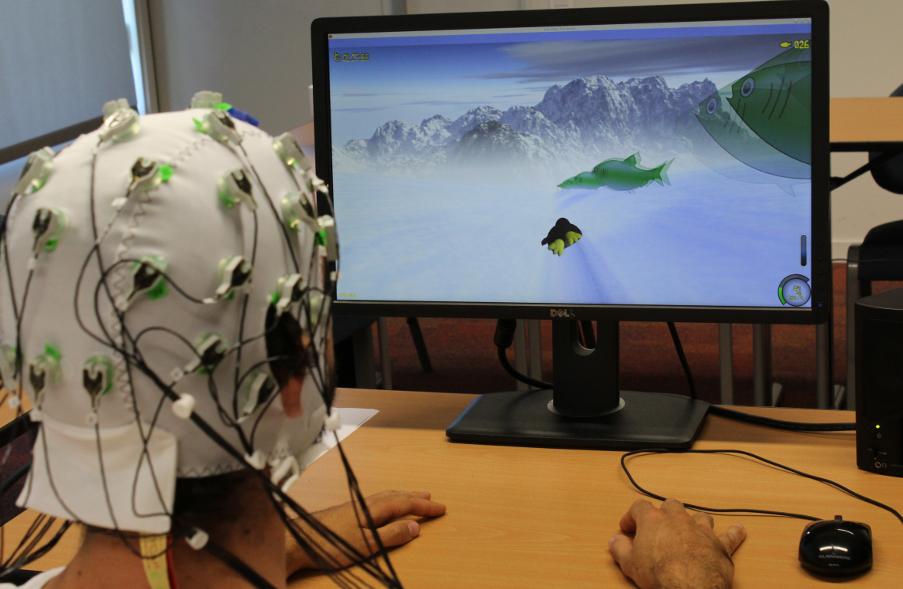}
    \caption{\small{Participant using MI commands to play TuxRacer, e.g. imagining right hand movement to catch fish on the right.}~\label{fig:tux_score}}
\end{figure}

\emph{Manipulating Flow:}	In order to fulfill Flow theory requirements, we considered the following:
\begin{itemize}

\item \textbf{An immersive and ludic environment}, here the \textbf{TuxRacer} video game was adjusted for a 2-class Motor Imagery (MI) BCI. The game depicts a ski course, in which a virtual penguin, Tux -- controlled by the player -- slides through various slopes and has to catch as much fish as possible. With the BCI adjustments, Tux was maneuvered with kinesthetic imagination of either left or right hand, see Figure \ref{fig:tux_score}.

\item \textbf{The adaptation} of the feedback bias, i.e. users were made to believe they performed differently from what they really did, in order to be in the flow state. If they had poor performances they were positively biased to a higher degree than if they had fairly good performances. However, when the performances were too good, then the users were slightly negatively biased, so that the task would not seem too easy. This was achieved by adaptively increasing or decreasing the classifier output, i.e. the decoding of MI commands would seem different from what it was in reality.

\item \textbf{Asynchronous music} consisted of 3 songs with medium tempo (120-160 beats per min), played in the background during the BCI task. 15 persons voted on social media for songs which would motivate them while playing TuxRacer. The selected songs are "Epic" by Alexey Anisimov (113s), "Confident \& Successful" by MFYM (168s) and "Acoustic Corporation" by OAP (132s), all available on Jamendo\footnote{https://www.jamendo.com/}.

\item \textbf{Clear goals with immediate audio and visual feedback}, i.e. to collect maximum points by manipulating Tux to move either left or right to catch fish.
The feedback is clear -- once caught, the fish disappears with a brief audio stimuli stressing that the target was reached.
\end{itemize}

\hspace{0.5cm}\emph{Experimental design:} We created a 2 (\textit{adapt} \textit{vs} \textit{no-adapt}) by 2 (\textit{music} \textit{vs} \textit{no-music}) mixed factorial design, i.e. a between-subject adaptation factor, and a within-subject background music factor.

\begin{figure}[!htbp]
    \centering
    \includegraphics[width=1\columnwidth]{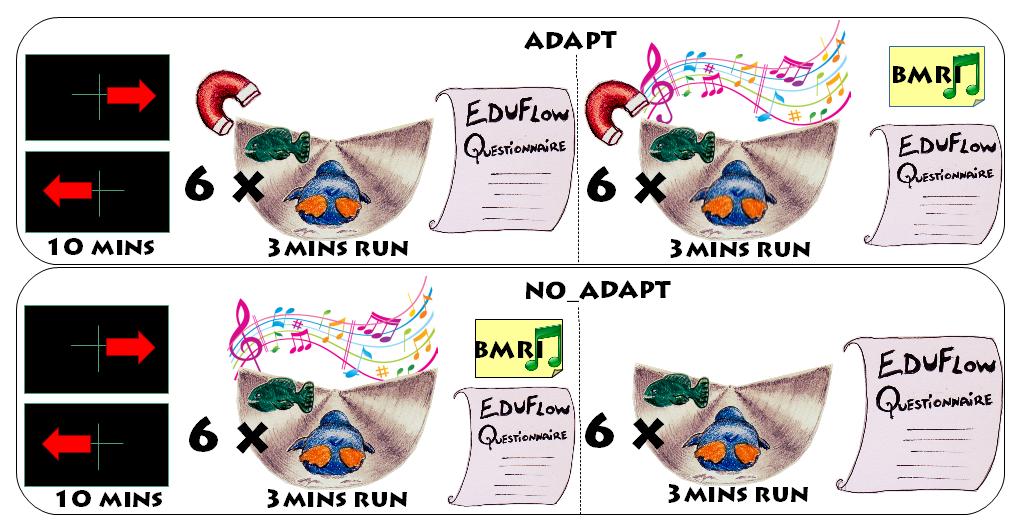}
    \caption{\small{The experiment started with \textasciitilde 10min calibration \cite{PfurtschellerMotorCommunication}, followed by 2 conditions: either with or without music -- 6 runs of 3 minutes per condition. The adapt group received an adapted (biased) feedback, contrary to the no-adapt group. Adaptation is symbolized by the magnet. Both groups were asked to fill EduFlow2 \cite{Heutte2016} questionnaires for the flow state assessment and BMRI \cite{Karageorghis1999} questionnaire for investigating the quality of music.}~\label{fig:protocool}}
\end{figure}

\hspace{0.5cm}\emph{Protocol:} 28 healthy subjects, naive to BCI, participated in the \textasciitilde2 hour-long experiment (5 women, mean age: 25.23 years, SD: 2.98). The first 30 minutes consisted of (i) signing a consent form, (ii) installing of a 32 channels Brain Product LiveAmp EEG, (iii) instructions given to the user and preparation, (iv) \textasciitilde10 minutes system calibration (40 trials of 7s) with the standard 2-class MI BCI (left/right hand) Graz protocol \cite{PfurtschellerMotorCommunication}. In the Graz protocol, the user was presented with arrows indicating the left or right side, to instruct the participant to imagine a left or right hand movement. Afterwards, each participant took part in 2 counterbalanced conditions of \textasciitilde 20mins each with TuxRacer, (a) with and (b) without background music. 3 songs were repeated to accompany the music condition of 6 runs (1 song per 2 runs). Each condition comprised of $6\times 3$min-runs, with 22 trials per run (11 for left and 11 for right hand, in random order), see Figure \ref{fig:protocool}. Each trial consisted in performing left/right hand MI to move Tux in order to catch fish on the left/right of the ski course, respectively. There were 7 closely arranged fish per trial, to be caught within 3 seconds. During 5-second long breaks between trials, the BCI controls were disabled so that Tux would return in a neutral position (center on the ski course) and participants could rest. The study was approved by the Inria ethics committee, COERLE (Comité opérationnel d'évaluation des risques légaux et éthiques).

\hspace{0.5cm}\emph{Questionnaires:} Prior to the experiment, a Swedish Flow Proneness Questionnaire (SFPQ) \cite{Ullen2011} was sent to subjects to fill in at home. This 5 points Likert scale questionnaire measures flow proneness -- flow as a person's trait. To estimate to which extent users were in the state of flow, they were asked to fill in the EduFlow2 questionnaire \cite{Heutte2016} after each condition (\textit{music} or \textit{no-music}). The EduFlow2 measures flow state through 4 dimensions: cognitive control, immersion, selflessness and autotelism -- a self rewarding experience. To have a measure of the quality and motivation of the selected music, the participants also filled a dedicated questionnaire, the Brunel Music Rating Inventory (BMRI) \cite{Karageorghis1999}.

\hspace{0.5cm}\emph{Signal processing}: The acquired EEG was band-bass filtered with a Butterworth temporal filter between 8 and 30Hz. We computed the band power using a 1s time window sliding every 1/16th s. We used a set of Common Spatial Patterns (CSP) spatial filters to reduce the 32 original channels down to 6 "virtual" channels that maximize the differences between the two class motor imagery \cite{Ramoser2000}. A probabilistic SVM (Support Vector Machine) with a linear kernel was used to classify the data between left and right classes (regularization parameter C = 1). That way, the output of the SVM between 0 and 1, indicated a class recognized with a certain degree of confidence, e.g. 1 means that the right- hand class was recognized with high confidence.

\hspace{0.5cm}\emph{Performances}:	The online performance corresponds to the peak accuracy of the classifier that controlled the video game, i.e. the highest classification accuracy over all trials' time windows. The offline performance was computed afterwards with a 4-folds cross validation, i.e. regarding only the data recorded during the interaction with the video game for training and testing. In other words, data recorded during the Graz protocol was not used to compute offline performances. We used a LDA (Linear Discriminant Analysis) for the offline classification, since it is less computationally demanding. Both for online and offline analyses, one accuracy score was computed over the music / no\_music condition (i.e. 6 runs of 22 trials).

\hspace{0.5cm}\emph{Game controls}: The TuxRacer game was controlled \textit{via} a virtual joystick. When a right hand movement was recognized (SVM output of 1) the virtual joystick was tilted toward the right at its maximum angle, 45 degrees. Inversely, when a left hand movement was recognized (SVM output of 0), the virtual joystick was tilted 45 degrees to the left. Between 0 and 1, the values of the virtual joystick were mapped linearly (from minus 45 degrees to 45 degrees). Thanks to this simple virtual joystick, we did not need to modify the usual input commands to the complex BCI ones in the game. Basically, the virtual joystick can act as or replace the usual computer controls, such as keys on the keyboard. Our freely available source code\footnote{https://github.com/conphyture/LSL2joy} could be used to control any (linux) joystick-based game with a BCI.

\hspace{0.5cm}\emph{Game modifications}: We designed the BCI TuxRacer game so that its timing and structure mirror that of the Graz motor imagery BCI protocol \cite{PfurtschellerMotorCommunication}, but in an immersive and motivating environment. We modified the shape of the terrain, curving it alike a bobsleigh course. Consequently, by the force of gravity, Tux would slide back to the middle of the screen between trials, when the commands were deactivated. Between trials, Tux would still be skiing towards the following trial with constant speed, enabling the users to see the next fish.
We fixed the position of the fish on the ski course edges, so that the targets were equidistant from Tux at the beginning of each trial, i.e. same distance from the center of the ski course. The reason for this is to enable the user to provide equivalent potential effort for both MI classes (left/right hand). By assuring a constant speed for Tux, a race (run) always lasted 3 minutes.

\hspace{0.5cm}\emph{Game adaptation}: The \textit{no\_adapt} group was the first to participate in the experiment. Thanks to the fact that there was a correlation between the user's flow state and the performances in the control group, we empirically calculated a performance level (classifier accuracy) for which users felt most in flow. We used that value as an attractor or a quasi-flow value to lure Tux in. At each instant ($1/16^{th}$sec sliding window) we would retrieve the classifier output and add to it a value which would push Tux a half way towards our attractor. This value was determined intuitively from Flow theory, to keep the difficulty in the "golden middle". Consequently, when user performances were very poor, Tux was boosted to a higher extent towards the attractor, i.e. in this case users were helped (positively biased) more than when their performances were fairly good. However, when the performances were too good, the perceived performances were deteriorated (feedback was negatively biased). The "flow" function would then be: $f(s_i) = s_i + \frac{(a-s_i)}{2} , s_i \in [-1,1] $, where $s_i$ stands for user skill, which is given by the classifier output and scaled to ease the computation (-1 for left, 1 for right), for all instants $i$ within 22 trials, ($i =1,..16Hz\times 66s$). Finally, $a = 0.79$ denotes the attractor for the right class ($a = -0.79$ for left class).

\section*{Results}

The normal distribution of all the data was verified using a Shapiro-Wilk normality test.

\begin{figure}[!htbp]
    \centering
    \includegraphics[width=0.9\columnwidth]{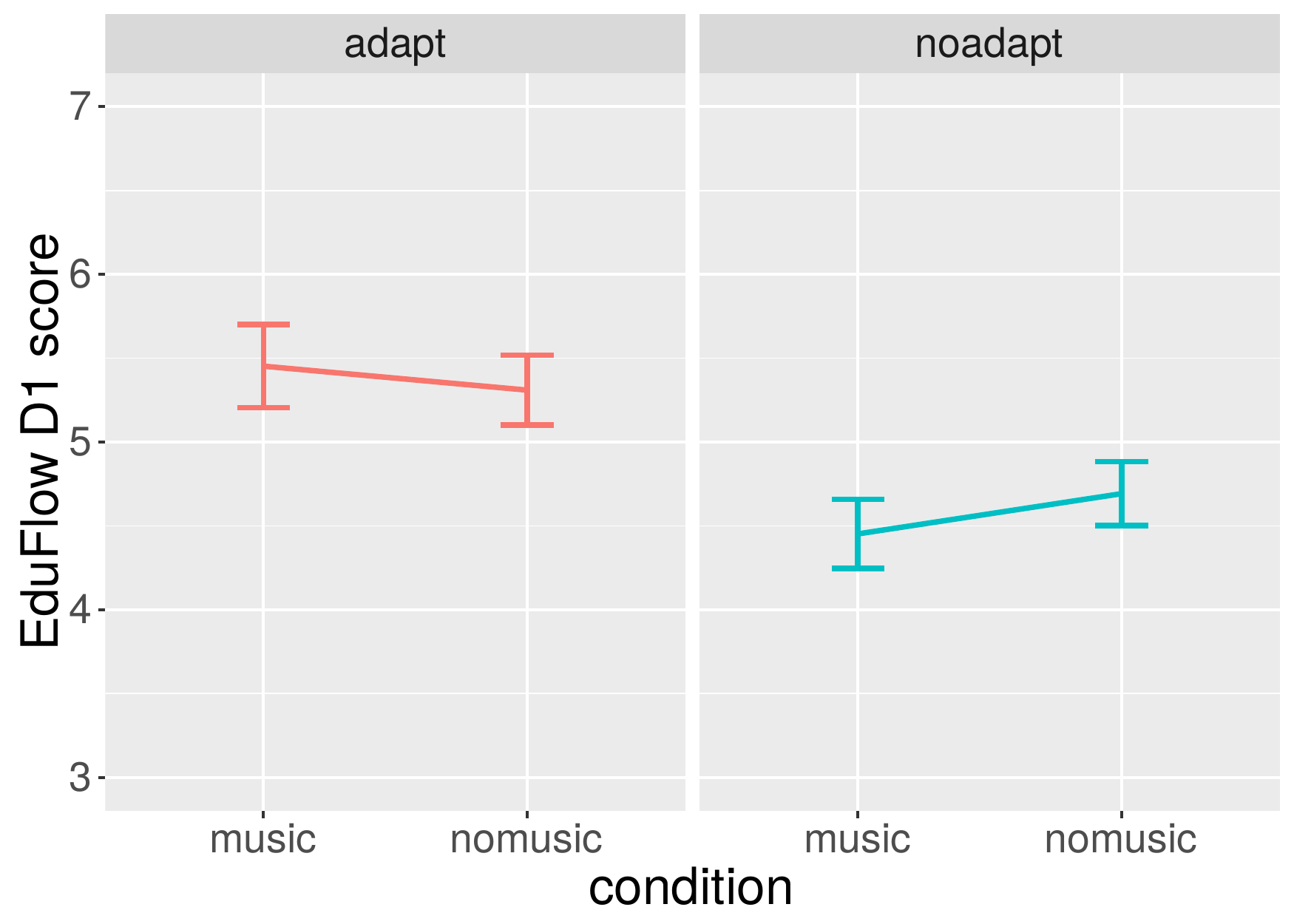}
    \caption{\small {The EduFlow2 score (7 Likert scale) for the first dimension (cognitive control) depending on the between-subject factor \textit{adapt} and on the within-subject factor \textit{music}. Users were in higher cognitive control in the adapt condition (left).}~\label{fig:eduflow}}
\end{figure}

\hspace{0.5cm}\emph{Flow-factor's influence on EduFlow2}: We tested the effects of our mixed factorial design on each of the 4 dimensions measured by the EduFlow2 questionnaires using a Markov Chain Monte Carlo (MCMC) method \cite{Hadfield2010b}. The MCMC showed a significant difference between \emph{adapt} and \emph{no-adapt} along the $1^{st}$ dimension (p < 0.01). Participants in the adapt group reported higher cognitive control (mean: 5.38, SD: 0.84) compared to the no-adapt group (mean: 4.49, SD: 0.83), see Figure \ref{fig:eduflow}.

There was no significant difference between groups regarding their flow trait, measured with SFPQ (1-way ANOVA, p = 0.25). ANCOVA tests showed that it was not a confounding factor for neither EduFlow2 nor performances.

There was no difference between groups regarding BMRI (1-way ANOVA, p = 0.53).  Mean score:  15.80, SD: 4.14 -- maximum score with the questionnaire we distributed: 25.3. There was no correlation between BMRI scores and flow (p = 0.54) nor with user performance, online (p = 0.78) or offline (p = 0.20). 

\begin{figure}[!htbp]
    \centering
    \includegraphics[width=0.9\columnwidth]{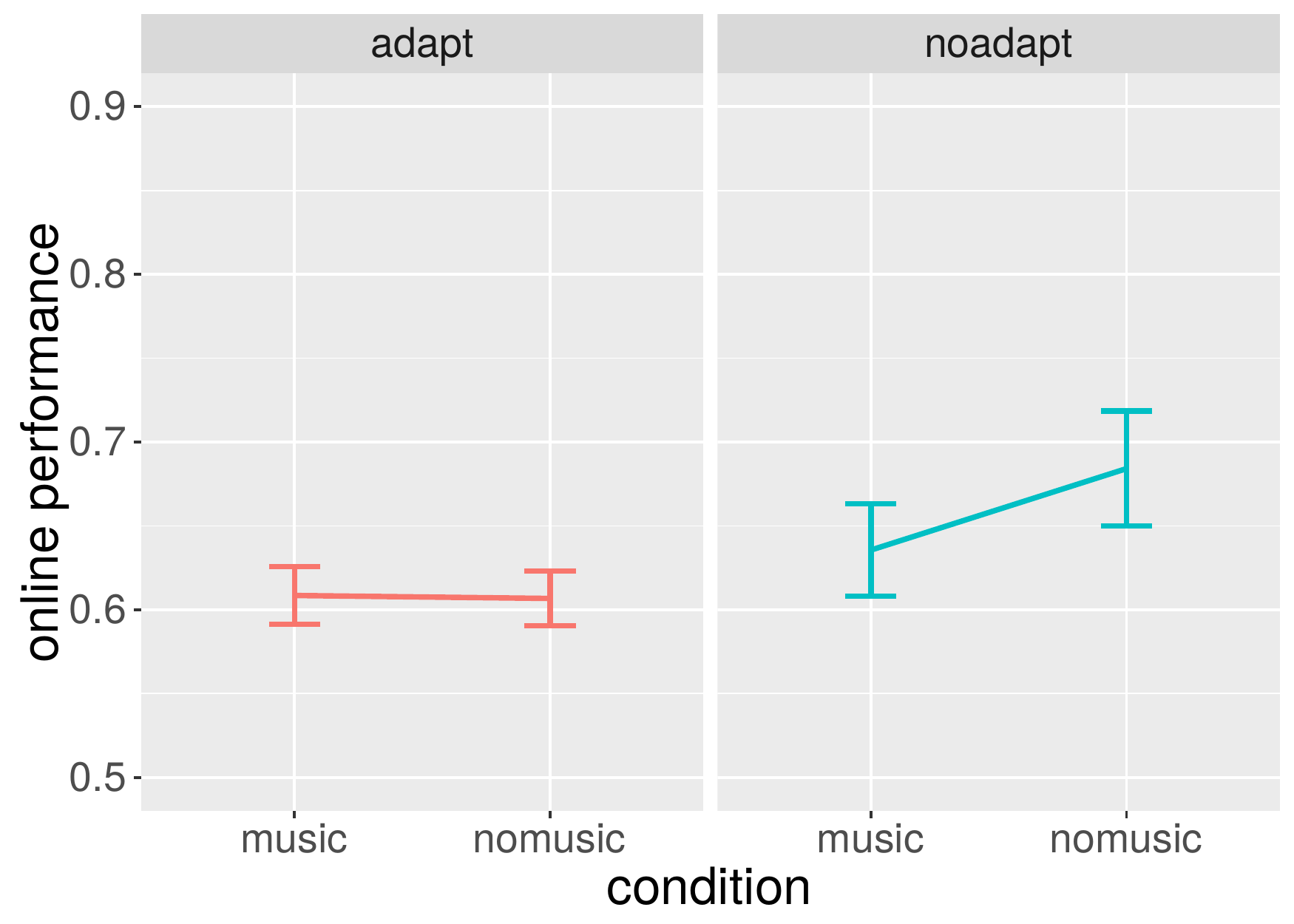}
    \caption{\small {The peak performance during the completion of the video game depending on the between-subject factor \textit{adapt} and on the within-subject factor \textit{music}. In the no\_adapt condition (right), users had better online performances without music.}~\label{fig:perfs}}
\end{figure}

\hspace{0.5cm}\emph{Flow-factor's influence on performance}: The question whether our conditions could directly improve the online performances was tested with a 2-way ANOVA. There was a significant interaction between music and adaptation (\textit{p<0.05}). Music had a significant effect on the online performance (mean with music: 0.62, SD: 0.09, mean with no\_music: 0.65, SD: 0.11, \textit{p<0.05}) but adaptation had not (\textit{p=0.08}), see Figure \ref{fig:perfs}. A post-hoc Tukey analysis reveals that the one significant interaction occurs in the no\_adapt condition, between music (mean: 0.64, SD: 0.11) and no\_music (mean: 0.68, SD: 0.13) (\textit{p<0.001}).

\begin{figure}[!htbp]
    \centering
    \includegraphics[width=0.9\columnwidth]{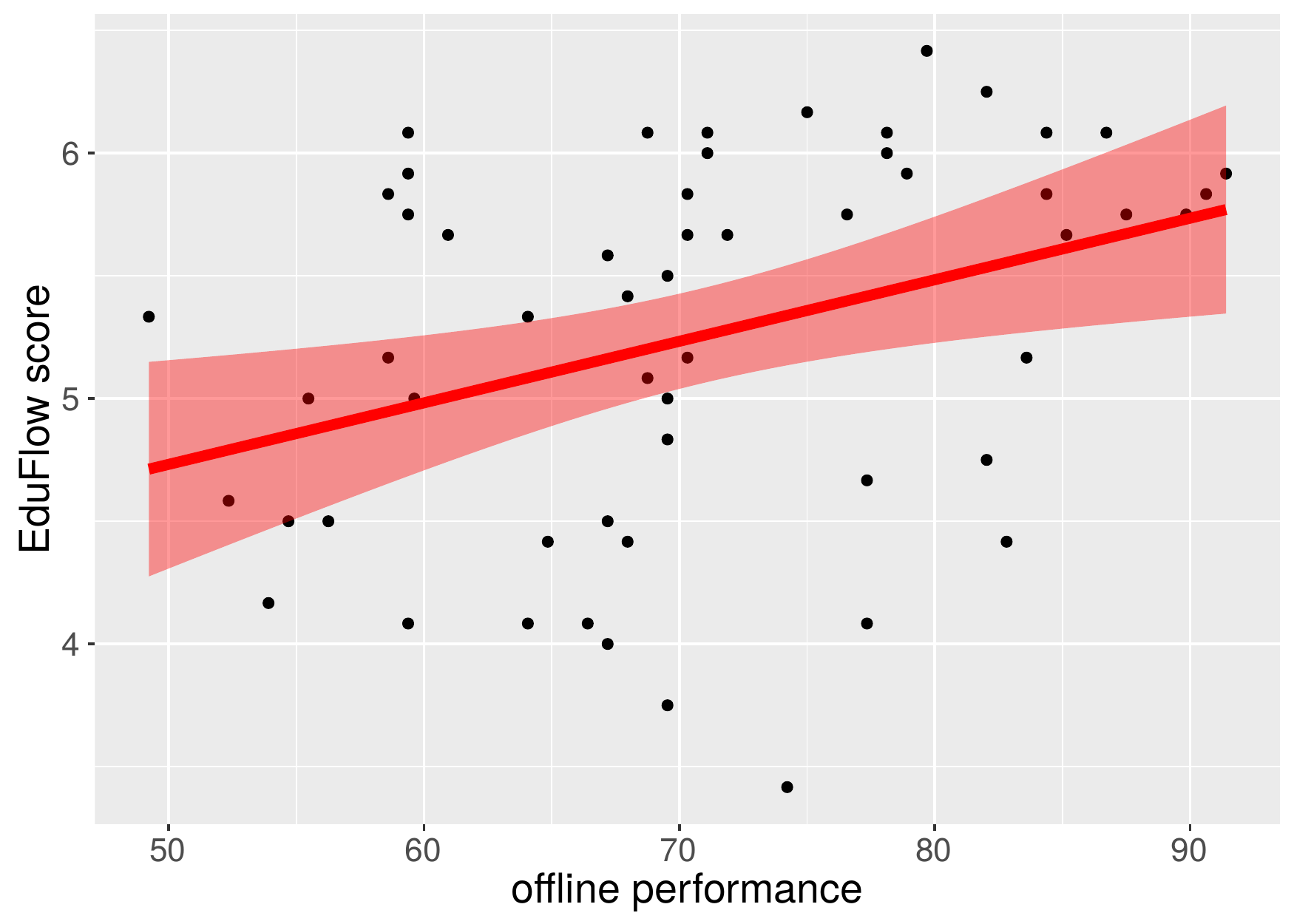}
    \caption{\small {Positive correlation between EduFLow2 scores (mean of 4D) and offline performance (Pearson coef:0.35, \textit{p<0.01}).}~\label{fig:correlation}}
\end{figure}

\hspace{0.5cm}\emph{Correlation between EduFlow2 \& performance}: 

There was no correlation between flow (mean of all the EduFlow2 dimensions) and online performance (\textit{p=0.12}), however there was a positive correlation between flow and offline performance (Pearson coefficient: 0.35, \textit{p<0.01}), see Figure \ref{fig:correlation}. More precisely, offline performances are significantly correlated with two dimensions of flow: the $2^{nd}$ -- immersion (p<0.01, Pearson coefficient: 0.38) and the $4^{th}$ -- autotelism, (p<0.05, Pearson coefficient: 0.32). We corrected the p-values for multiple comparisons with false discovery rate \cite{Noble2009}.

\section*{Discussion}

\textbf{H1. validated}: Adapting the task difficulty to users skill improved one dimension of flow state, cognitive control. People who faced a challenge better suited to their skill felt more in control. Thus, taking into account user's predispositions could lead to a greater user experience.

\textbf{H2. in contradiction}: Not only the presence of a background music had no effect on flow, but it deteriorated the online performance. Therefore, this result contradicts our second hypothesis.

\hspace{0.5cm}\emph{Music pace mismatch}. As opposed to what we expected, we could not directly improve performance by manipulating the flow factors we chose (adaptation and music). The latter could be explained by the songs we chose, since the motivational qualities of the music (measured with the BMRI questionnaire) were not very high and not correlated to any dimension of flow. Instead of picking those songs from the public domain, users may have been more motivated should they have chosen their own music. The decrease of performance in the music condition might come from the mismatch between the rhythm of the music and the pace of the game, i.e. with the pace of the imagined hands movements. Indeed, some users shared informally that they were imagining playing their musical instrument as MI commands and that the songs further disturbed their pace.

\hspace{0.5cm}\emph{Different training environment.} There was no correlation between flow state (EduFlow2) and online performances. That could be due to the differences between the calibration environment (Graz protocol) and the game, e.g. the first being minimalistic and the latter a 3D video-game. Moreover, as the calibration was done without music, maybe the performances online were better without it because the EEG signals might have changed, therefore the classifier could not recognize them anymore.

\hspace{0.5cm}\emph{Flow increases with performances.} These later assumptions are strengthened by the fact that there was a positive correlation between flow and the offline performances, when only game data was taken into consideration. The state of flow was then positively correlated with users' performance: the feeling of immersion and the autotelic experience (i.e. the completion of the task was self-rewarding) increased with the offline performance. Hence, not only encouraging a state of flow would produce BCIs more pleasing to the users, but it might also benefit the accuracy of the system. We still have to identify the direction of the correlation though: does flow state increases performances or do good performances increase flow state?

Overall, the discrepancy in our results could stress that flow is a complex phenomenon, and however beneficial to obtaining better BCI, the emerging interaction between its components should be more thoroughly investigated. 

\section*{Conclusion}

By investigating means to improve BCI user performance and usability through instructional design theories, we came across the Flow Theory. This theory, which describes an optimal user state, showed to improve performances in many fields. We hypothesized that the state of flow could benefit BCIs. In a MI BCI task, we manipulated flow by adapting the perceived difficulty and by adding a background music. We used an immersive environment, a 3D video game, TuxRacer (the modification can be found online \footnote{https://github.com/jelenaLis/tux-modifs)}). 

Our main findings show that the adaptation increases one of the dimensions of flow -- cognitive control, and that user's offline performances are positively correlated with flow. In the future we could attempt to better suit the adaptation of the task to the users: it could be biased adaptively over time, across several sessions, following the progress of the user. We could also try to account for the amount of effort that the user puts into the completion of the task in order to better comprehend such complex phenomena. For example, measuring workload could facilitate the assessment of the challenge that users are facing and computationally predict the state of flow\cite{Bulitko2012FlowModel}.

According to the literature, we chose asynchronous music with medium tempo to follow the BCI task. Unexpectedly, the background music impeded the performances of the user. This result stresses the importance of the choice of music to accompany a task. One explanation could lie in the very BCI paradigm we chose. Indeed, a motor imagery task might share similarities with actual physical activity, where it had been shown that \textit{synchronous} music could effectively stimulate the sensory-motor cortex\cite{Hardy2013}. Hence, a future work would consist in synchronizing music to game's cues (e.g. trials sequences) or to user's motor imagery pace. Such music, generated in real time, might enhance the flow state and intrinsic motivation. Concurrently, we should verify if the user is musically educated, as in some cases users imagined playing instruments as MI commands, and because musicians elicit different brain activity in motor areas\cite{Luo2012}. 

Flow is not only a promising research direction to improve BCI systems, but it raises a new question: should we put all our efforts in favoring the machine accuracy, or rather the human subjective experience?

\smallbreak

\emph{Acknowledgements}: This work was supported by the French National Research Agency with the REBEL project and grant ANR-15-CE23-0013-01. The authors would like to thank Jean Heute for his help with the EduFlow questionaire.

\setlength{\bibitemsep}{0pt}
\printbibliography

\end{document}